\documentclass[10pt,sigconf,letterpaper,nonacm]{acmart}

\usepackage[english]{babel}

\usepackage{enumitem}
\usepackage{cleveref}
\usepackage{tabularx}
\usepackage{siunitx}
\usepackage{ragged2e}  \newcolumntype{Y}{>{\RaggedRight\arraybackslash}X}

\renewcommand\footnotetextcopyrightpermission[1]{} \setcopyright{none}

\acmDOI{}

\acmISBN{}

\acmPrice{}

\usepackage{balance}

\usepackage{subcaption} 
 \usepackage{graphicx}
\usepackage[export]{adjustbox}
\usepackage{amsmath}
\usepackage{multicol,multirow}
\usepackage{epsf,psfrag,pstricks}
\usepackage{mdwlist}
\usepackage{rotating}
\usepackage{xspace}
\usepackage{fancyvrb}
\usepackage{tcolorbox}
\usepackage{url}
\usepackage{breakurl}

\usepackage{capt-of}
\usepackage{flushend}
\usepackage{soul}

\DeclareSIUnit[per-mode=symbol]\bps{\bit \per \second}
\DeclareSIUnit[per-mode=symbol]\kbps{\kilo\bps}
\DeclareSIUnit[per-mode=symbol]\Mbps{\mega\bps}
\DeclareSIUnit[per-mode=symbol]\Gbps{\giga\bps}
\DeclareSIUnit{\nothing}{\relax}

\newcommand{\ie}{i.e., \@}
\newcommand{\eg}{e.g., \@}

\newcommand{\cf}{cf. \@}

\newcommand{\etal}{et al.\xspace}

\graphicspath{{figures/}}

\begin{document}

\title{Geoflaw: Geofeed Errors, Inconsistencies, and Flaws}

\author{Oliver Gasser}
\affiliation{\institution{IPinfo}
\country{}
}
\author{Robert Beverly}
\affiliation{\institution{San Diego State University}
\country{}
}

\begin{abstract}
Geofeeds allow IP address resource owners to self-publish the
geolocation of their prefixes, and adoption has grown substantially:
as of November 2026, more than 5,000 geofeeds cover 1.6M prefixes.  We measure
geofeed errors, inconsistencies, and flaws by collecting geofeed
references from the WHOIS of all five RIRs weekly over three months
and downloading every referenced geofeed.  The share of WHOIS objects
complying with RFC~9632, which specifies how inetnum objects reference
geofeeds, drops from 90\% to 61\%, driven by a single operator's
objects with multiple geofeed references.  Each week, 12--20\% of
geofeed URLs cannot be downloaded, and one in five downloaded geofeeds
violates RFC~8805, which defines the geofeed CSV format.  Moreover, 17\% of geofeeds list IPv4 prefixes whose
WHOIS object references a different geofeed.  Finally, 281k prefixes appear
in more than one geofeed, 3.8k of them with conflicting locations, and
for 26\% of these conflicts no geofeed is authoritative.  We publish
our results on a public dashboard, which one large CDN has already
used to fix its geofeeds.

\end{abstract}

\maketitle

\section{Introduction}

IP geolocation maps IP addresses to physical locations.  IP
geolocation supports critical Internet services by providing \eg
access control, content customization, network management, targeted
advertising, fraud prevention, and policy enforcement.  This work
focuses on ``geofeeds.'' 

Geofeeds are a standard CSV format for representing IP prefix
geolocations~\cite{rfc8805}, while geofeed URLs are now standardized as a
RPSL field for WHOIS inetnum objects~\cite{rfc9632}.   Geofeeds are an
increasingly popular means for providers to self-publish and
self-report geolocation information; the geolocatemuch project reports 713,460
IP prefixes with geofeed locations on November 4, 2026~\cite{geolocatemuch}.
However, our independent geofeed gathering and processing methodology
finds 1,637,955 prefixes on the same date.  This large discrepancy
(2.3 times as many prefixes) motivates our study of errors and
inconsistencies within RPSL entries, geofeed hosting, and the
geofeeds themselves.  These
validation issues significantly impact the ingestion and use
of geofeeds by geolocation providers, often unbeknownst to the
networks publishing geofeeds.  Our contributions include:
\begin{itemize}[leftmargin=*, topsep=2pt, itemsep=1pt, parsep=0pt]
 \item \textbf{Longitudinal WHOIS compliance analysis
       (\S\ref{sec:methodology}):} Across three months of weekly
       snapshots from all five RIRs, the share of WHOIS geofeed
       references that comply with RFC~9632
       drops from 90\% to 61\% due to a single operator,
       and about 30\% of LACNIC's objects reference geofeeds via HTTP
       instead of HTTPS.
 \item \textbf{Download, format, and coverage errors
       (\S\ref{sec:methodology}):} Each week, 12--20\% of geofeed URLs
       cannot be downloaded, one in five geofeeds violates RFC~8805,
       and 17\% 
list prefixes whose WHOIS object
       references a different geofeed.
 \item \textbf{Inconsistencies between geofeeds
       (\S\ref{sec:inconsistencies}):} Of the 1.6M geofeed prefixes,
       281k (17\%) appear in multiple geofeeds, 3.8k of them with
       conflicting locations, and 26\% of these conflicts lack an
       authoritative geofeed.  We also identify cases that RFC~9632
       does not resolve, such as an inetnum object referencing
       multiple contradicting geofeeds.
 \item \textbf{Geoflaw Geofeed Observatory
       Dashboard~\cite{geoflawweb}:} An interactive and queryable
       interface to our data for the network operator community, which
       one large CDN already used to fix its geofeeds.
\end{itemize}
 \section{Background}

While not originally part of the Internet infrastructure, a large body
of research developed \emph{inferential} IP geolocation techniques
such as Constraint Based Geolocation (CBG)~\cite{gueye2004constraint,du2020ripe}.
Augmenting CBG are databases such as the WHOIS and
DNS~\cite{dan2021ip}.
Unfortunately, registration databases can be inaccurate,
coarse-grained, and present conflicting
information~\cite{beverly2026whereis,darwich2023replication,poese2011ip}, 
while the accuracy of
inferential techniques depends strongly on the placement of
measurement nodes~\cite{shavitt2011geolocation,katz2006towards,hu2012towards}.

To address the need for reliable geolocations,
RFC~8805~\cite{rfc8805} introduced ``geofeeds:'' a simple UTF-8~\cite{rfc3629} CSV delimited
format for representing IP prefix geolocations.  RFC~9632~\cite{rfc9632}
then specifies a mechanism for IP address owners to
self-publish geofeed information for their resources by adding a new
element to the Routing Policy Specification Language
(RPSL)~\cite{rfc2622}.  The owner of an IP prefix can add a geofeed
URL to their IPv4 or IPv6 WHOIS inetnum object and create a publicly
downloadable file available at that URL that encodes ground-truth 
geolocations.
Conversely, consumers, \eg 
geolocation providers, can lookup an inetnum object within the WHOIS,
find the geofeed URL, download the geofeed, and obtain geolocations
for subnets within that prefix as set by the provider.  
RFC~9877~\cite{rfc9877}
additionally allows RDAP servers to return geofeed URLs for IP
network objects.

Since their introduction in 2020, geofeeds have seen steady adoption
and increasing use.  Fainchtein \etal~\cite{fainchtein2024you}
found 34.3M IPv4 addresses in October 2023;
data from our collection
(\S\ref{sec:methodology}) finds over 5,000 distinct geofeeds
covering over 1.6M IP prefixes and nearly 335M IPv4 addresses -- a
ten-fold increase over prior reported figures.  Recent activity with the IETF
identifies several limitations and potential improvements for
geofeeds~\cite{iabworkshop}; position papers at the corresponding
IAB workshop report malformed geofeed entries, false granularity
claims, and stale data~\cite{gasser2025challenges}. 
Prior research verified
self-reported geolocations against CBG~\cite{livadariu2024geofeeds}
and validated Starlink's geofeeds against anycast and traceroute
measurements~\cite{simon2026first}.
Closest to our work, Desai \etal found 7.8\% of geofeed URLs
to be inaccessible and 10.5\% of geofeed lines to be malformed
in early 2024~\cite{desai2025geofeed}, Tariq \etal could not
retrieve 9.5\% of 6,064 geofeed URLs and found RFC~8805 violations
in 80 geofeeds in September 2026~\cite{tariq2026where}, and
Mok \etal report non-standard geofeed references in the WHOIS and
duplicate entries across geofeeds in ongoing longitudinal
work~\cite{mok2026enabling}.
Complementing these studies, we examine the real-world
adherence to the geofeed RFCs weekly over three months across all
five RIRs, from the geofeed references in the WHOIS over the
geofeed downloads to the geofeeds' content, coverage, and conflicts,
and find significant opportunity to improve the
deployment practice of geofeeds.

\section{Downloading and Format Errors}
\label{sec:methodology}

We obtain the bulk WHOIS data from all five RIRs every Wednesday from
August 12 to November 4, 2026 (13 snapshots).  The WHOIS data is processed to
identify objects containing either a ``geofeed'' attribute, or a
``remarks'' attribute with the ``Geofeed'' token; this accommodates
different RIR and operational practice consistent with~\cite{rfc9632}.
After finding geofeeds, we attempt to download them via
their geofeed URL.  We perform all downloads on the same day as the
bulk WHOIS, and perform the fetch from a well-connected machine within
our institution.  The downloader randomizes the order of download, 
fetches sequentially, and fetches any given URL at most once to minimize 
potential impact on remote systems.  If the downloader experiences a 
transient failure, \eg DNS or timeout, it retries up to three times.

The same geofeed can be referenced by multiple different URLs, \eg due to HTTP redirection, HTTP vs. HTTPS, or mirrors.
Counting unique URLs would therefore lead to the same geofeed being counted multiple times.
Thus, we identify geofeeds by the SHA-256 checksum of their content.

In the November 4 snapshot, we successfully download geofeeds from 5,891 URLs, of which 5,702 contain geofeed entries.
When using the SHA-256 identifier on the content of the geofeed files, we end up with 5,058 unique geofeed files, 123 of these are referenced by more than one URL.
Two geofeed files account for most of these duplicates, \ie kinako.network with 256 URLs and hatsnet.io with 253.

\subsection{RFC~9632 Compliance}

\begin{figure}[!t]
\centering
\includegraphics[width=1.0\linewidth]{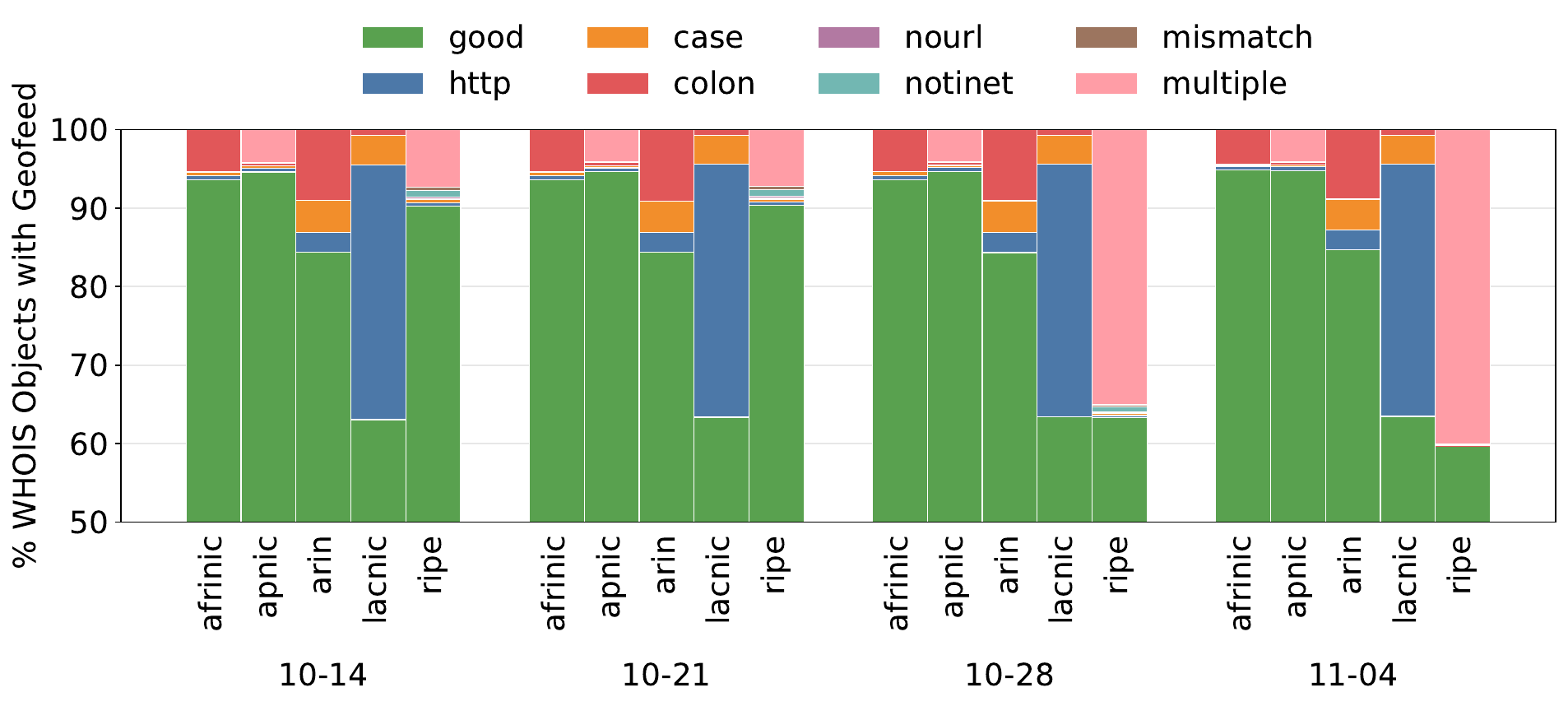}
\vspace{-7mm}
\caption{WHOIS object adherence to RFC~9632~\cite{rfc9632}
(last four of 13 weekly snapshots).}
\label{fig:rfc9632}
\end{figure}
\vspace{-2mm}

We first consider adherence to RFC~9632~\cite{rfc9632}, the specification for finding
Geofeeds within the RPSL.  In processing the bulk WHOIS data, we find
several common classes of deviation from the RFC-mandated behavior.
First, the ``Geofeed'' token following the ``remarks'' attribute often
has the wrong case, e.g., ``geofeed'', or adds an extraneous colon,
e.g., ``Geofeed:'' in violation of the RFC's case sensitivity and
formatting requirements.  Second, the URL following a ``Geofeed''
token or within a ``geofeed:'' attribute can be either malformed or
entirely missing.  Third, the object may contain multiple geofeed
attributes that point to different geofeed URLs.  Fourth, the geofeed
may be contained in an object other than an inetnum or inet6num, for
example within an aut-num object -- also a violation of the RFC.
Finally, while geofeeds must be published via HTTPS, many are not.

Figure~\ref{fig:rfc9632} shows the proportion of these errors
relative to the total objects with geofeeds over time.
We observe per-RIR differences, for instance LACNIC has a small
number of geofeeds listed without HTTPS for about 30\% of their
objects; this effect is exacerbated by the overall low-level
of geofeed adoption within LACNIC.  ARIN objects exhibit a
non-negligible fraction of objects with either casing or colon 
errors, while RIPE has the highest fraction of non-inetnum
objects with geofeeds.  Finally, we note a significant change
in multiple geofeed entries per object for RIPE starting with
our October 28 snapshot.  A deeper investigation reveals a 
single provider that added multiple geofeed attributes to 
nearly 800k objects on that date, demonstrating how individual
resource owners can significantly impact the ecosystem today.
Across all RIRs, the share of RFC-compliant objects drops from
around 90\% until October 21 to 61\% on November 4.

RFC~9632~\cite{rfc9632} also documents a mechanism to authenticate geofeeds using RPKI~\cite{rfc6480} signatures.
We check for RPKI signatures in all downloaded geofeeds but find none.

\subsection{Geofeed Download}

\begin{figure}[!t]
\centering
\includegraphics[width=1.0\linewidth]{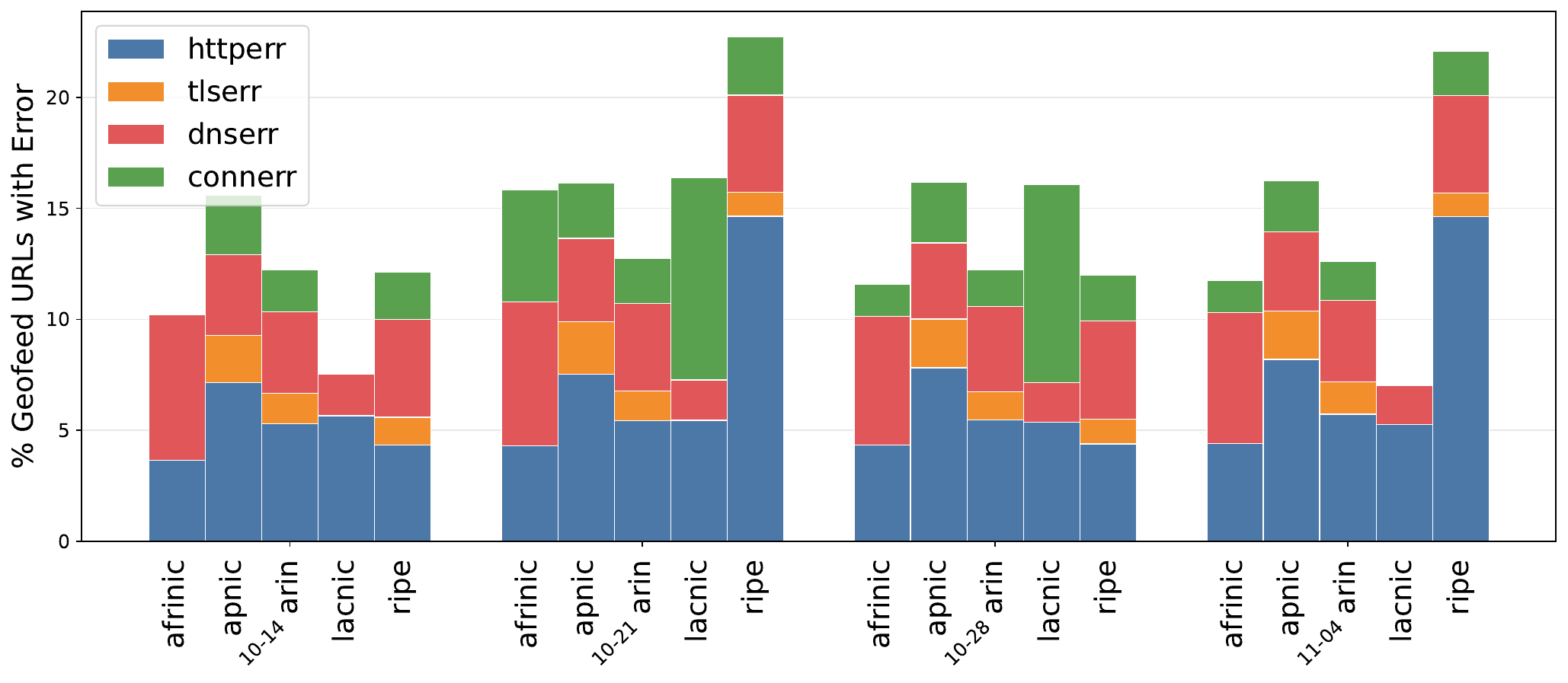}
\vspace{-4mm}
\caption{Geofeed download errors (last 4 snapshots).}
\label{fig:fetch}
\end{figure}
\vspace{-2mm}

The presence of a geofeed within a WHOIS object does not imply that
the URL exists, or that it can be successfully downloaded.
Figure~\ref{fig:fetch} displays the fraction of various download
errors, as a function of time and RIR.  Across all 13 snapshots, APNIC,
ARIN, and RIPE exhibit TLS error rates between approximately 1--8\%,
primarily due to expired or invalid certificates.  Similarly, the
fetcher is unable to resolve the name for 2--8\% of geofeeds.
A non-negligible number of geofeeds simply do not respond after DNS
resolution, or otherwise result in a connection error.  Finally, a
consistent fraction of requests result in an HTTP error.  We observe
15 different HTTP error codes, with 404 being the most common.  The
large change in HTTP errors for RIPE toward the end of our collection
is due to a single provider returning 502, bad gateway, errors for
many geofeeds (548 of the 553 URLs failing with a 502 on November 4).
In total, 12--20\% of the geofeed URLs cannot be downloaded
in each snapshot.  Overall, we observe that the ability of geolocation
providers or other consumers to obtain geofeed data varies
over time and RIR, but a significant fraction of published geofeeds
remain consistently unavailable.

\subsection{RFC~8805 Compliance}

RFC~8805~\cite{rfc8805} defines the data format for geolocation files, and provides
sample validation code, as well as test vectors.  We utilize the
``geofeed-validator'' Python library to assess the geofeeds we are
able to download~\cite{geofeedvalidator}.  Unfortunately, some 
in-the-wild geofeeds contain many thousands of
identical lines that cause this library to crash due to a quadratic
memory loop attempting to maintain all-pairs duplicates.  We fixed and
provided a patch to the authors~\cite{geofeedpatch}, and then ran all
geofeeds through this reference validation.
 
\begin{figure}[!t]
\centering
\includegraphics[width=0.9\linewidth]{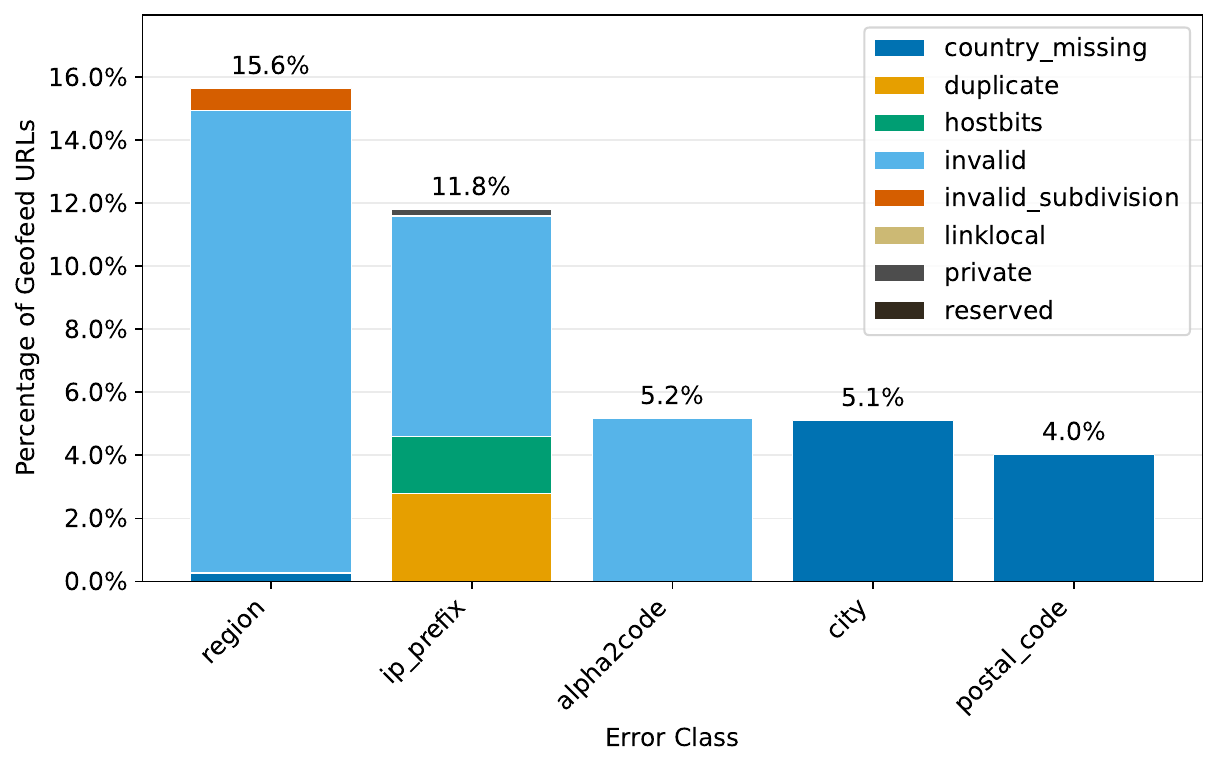}
\vspace{-4mm}
\caption{Adherence of geofeeds to RFC~8805~\cite{rfc8805} as 
determined by the geofeed-validator~\cite{geofeedvalidator} reference.}
\label{fig:rfc8805}
\end{figure}

Figure~\ref{fig:rfc8805} divides the geofeed validation errors into
five primary categories, corresponding to the primary error categories
from the reference validation script.  Within each category are the
specific errors.  We validate the file served by each geofeed URL.
In total, 1,221 URLs, \ie about one in five,
serve a file with at least one validation error.  The most common validation error is
invalid regions, either due to an invalid ISO~3166-2~\cite{iso3166-2} subdivision code,
or less commonly, the region not belonging to the country.  5.1\% of
the URLs serve an entry where the city is specified, but the
accompanying country is invalid or missing, while 4.0\% serve a
postcode but not country. 5.2\% of the URLs serve an entry with
an invalid ISO~3166-1~\cite{iso3166-1} alpha-2 code.  The second most common errors are
prefix errors, typically due to the file including an invalid
prefix.  Almost 2\% of the URLs serve a prefix where the host
bits are set~\cite{rfc4632}, while nearly 3\% serve a duplicate prefix.
\Cref{sec:inconsistencies} examines these duplicates in detail.

\subsection{Coverage}

Next, we consider the alignment of prefixes within geofeed
files to their corresponding covering inetnum objects within
the WHOIS -- we term this a prefix's ``coverage.''
The geofeed standard states that ``one MUST ignore
data outside the referring inetnum object's address range''~\cite{rfc9632}
in order to ensure that the data is under the control of
the operator.   

We first build a radix trie of IPv4 and IPv6 prefixes
inclusive of those inetnum and inet6num objects within
the bulk WHOIS that include a geofeed.  Each node in the
radix trie is the corresponding geofeed (SHA-256 hash of
its contents).  We then assign to every prefix within 
all downloaded geofeeds one of three coverage labels.
First, if a prefix within a geofeed $G$ belongs to (\ie is
a subnet of) a
WHOIS object that also contains a geofeed reference to $G$,
we term the prefix ``covered.'' If the corresponding object
has no geofeed reference, we say that it is ``not referenced.''
Finally, if the object has a geofeed attribute, but references
a different geofeed, we term the prefix ``not covered.'' 
Table~\ref{tab:labels} provides an example of each case.

Note that, a ``not covered'' prefix in a geofeed file could
be a mistake, or could represent an attempt to provide 
geolocation data for a prefix not under the control of the
prefix owner.  To validate our methodology, we publish 
both IPv4 and IPv6 prefixes within a geofeed, where that
geofeed belongs to inetnum other than the listed geofeed
prefixes.  Our methodology detects both of these ``not
covered'' canaries.

\begin{table}
\caption{A geofeed $G$ containing a prefix
covered by an inetnum object referencing $G$
is \emph{covered}.  If the inetnum object has no
geofeed reference, the prefix is \emph{not referenced}. Finally,
if the covering inetnum object references a geofeed other than $G$,
the prefix is \emph{not covered}.}
\vspace{-2mm}
\label{tab:labels}
\centering
\resizebox{1.0\columnwidth}{!}{
\begin{tabular}{|l|l|l|l|} \hline
\textbf{Geofeed}  &  \textbf{WHOIS}   & \textbf{WHOIS}   & \\
\textbf{$G$}      &  \textbf{inetnum} & \textbf{Geofeed} & \textbf{Label} \\
\midrule
\texttt{1.2.3.0/24} &  \texttt{1.2.3.0/20}    & $G$             &
Covered \\
\texttt{6.7.8.0/24} &  \texttt{6.7.8.0/20}    & none          &
Not referenced \\
\texttt{9.8.7.0/24} &  \texttt{9.8.7.0/20}    & $H$             &
Not covered \\
\bottomrule
\end{tabular}
}
\end{table}

\begin{figure}[!t]
\centering
\includegraphics[width=0.9\linewidth]{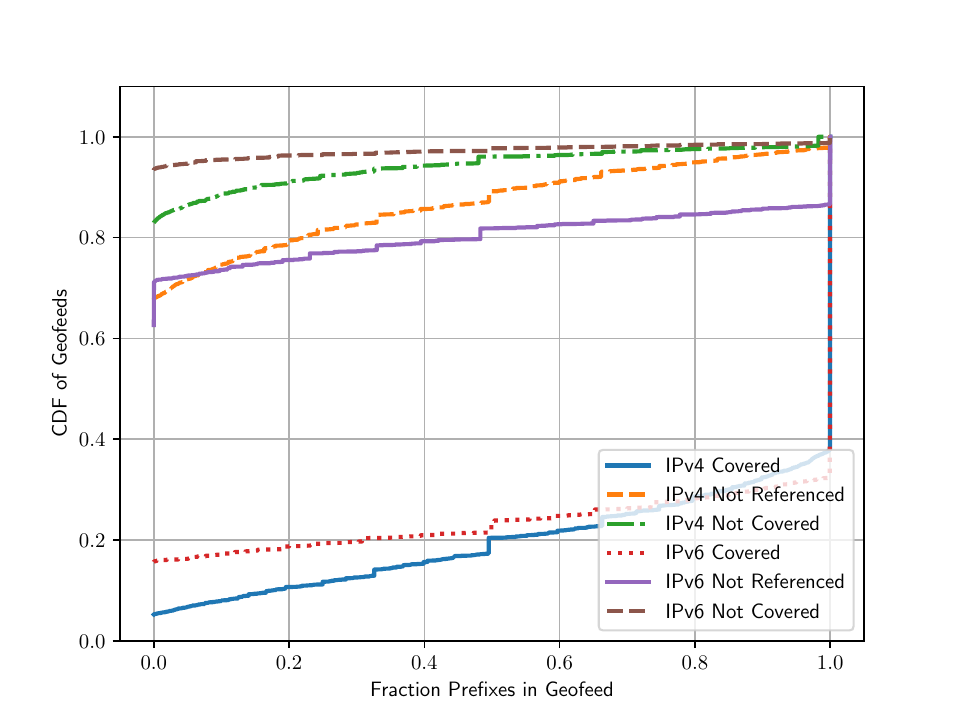}
\vspace{-4mm}
\caption{Prefix coverage (Table~\ref{tab:labels}) 
across geofeeds.}
\label{fig:coverage}
\end{figure}

For each collected geofeed, we determine the fraction of prefixes that
are in each coverage category, and also between IPv4 and IPv6 protocol
versions.  Figure~\ref{fig:coverage} displays the cumulative
distribution of geofeeds as a function of the fraction of prefixes
within each coverage category.

We observe that for $\sim$60\% of the geofeeds, all of the IPv4 prefixes are
covered, while 5\% of the geofeeds contain no covered IPv4 prefixes.
Further, 35\% of the geofeeds contain IPv4 prefixes that are not
referenced, while approximately 17\% of the geofeeds have some 
IPv4 prefixes that are not
covered.  Finally, we see that more geofeeds have no
covered IPv6 prefixes as compared to covered IPv4 prefixes.

 \section{Geofeed Entry Inconsistencies}
\label{sec:inconsistencies}

Next, we will look into inconsistencies of the geofeed files themselves by inspecting geofeed entries for the November 4, 2026 snapshot.
We check for prefixes with multiple entries in geofeed files, how often and how they disagree on locations, analyzing which entry is authoritative, and checking for inconsistencies across RIRs and countries.

\subsection{Duplicate Entries Within a Geofeed}
\label{sec:inconsistencies:sub:within}

According to the geofeed standard~\cite{rfc8805}, there should only be a single geofeed entry for a specific prefix.
We check if this is indeed the case, first within the same geofeed file and then across geofeed files, where we term prefixes listed by at least two geofeeds \emph{multi-entry prefixes}.

\begin{figure}[t]
  \centering
  \includegraphics[width=\columnwidth]{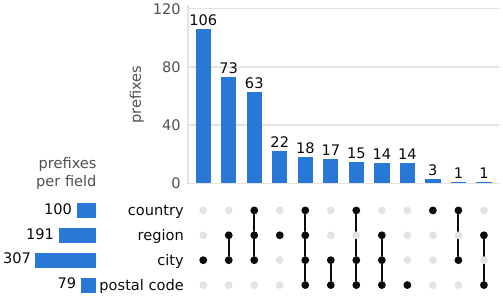}
  \caption{Fields differing between entries of prefixes listed more than once within the same geofeed.}
  \label{fig:duplicates_upset}
\end{figure}

We first look at duplicate entries within the same geofeed file.
In total we find 2662 duplicate entries in 186 geofeed files.
The vast majority (2315) of these duplicates are completely identical, \ie all geolocation fields for a prefix are the same.
For 347 duplicate prefix entries we see conflicting geolocation information in the same geofeed.
As shown in \Cref{fig:duplicates_upset} the most common differences are in the city field; the city and region fields; and the country, region, and city fields.

We investigate the characteristics of these duplicate differing entries and find that only 3 out of 294 city pairs are likely typos (Levenshtein distance~\cite{levenshtein1966binary} $\leq 2$).
50 of the 344 city pairs have one side empty, \ie one geofeed entry lists fine-grained city-level location while the other lists only coarse grained region- or country-level location.
We do also see countries changing in 100 prefixes, likely between PoP locations for hosting, VPN, and IP-leasing providers~\cite{du2024sublet} (\eg HK/SG, AU/HK, GB/US, HK/TW, DE/NL).

\subsection{Duplicate Entries Across Geofeeds}
\label{sec:inconsistencies:sub:across}

Next, we analyze duplicate geofeed entries across different geofeeds.
Of the 1.6M geofeed prefixes, 280.7k are listed in at least two distinct geofeeds.
Duplicate geofeed entries are almost twice as common for IPv4 (20.5\%) compared to IPv6 (10.7\%).
Two geofeed pairs stand out here:
Metronet's HTTP~\cite{metronetgeofeedhttp} and HTTPS~\cite{metronetgeofeedhttps} geofeeds with 160.6k prefixes and Cloudflare's local-ip-ranges.csv~\cite{cloudflarelocalipranges} and warp-egress-ip-ranges.csv~\cite{cloudflarewarpegress} with 88.1k prefixes.

\subsection{Conflicting Location Information Between Geofeeds}
\label{sec:inconsistencies:sub:conflicts}

\begin{figure}[t]
  \centering
  \includegraphics[width=\columnwidth]{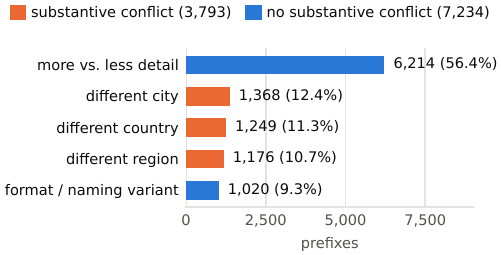}
  \caption{Prefixes listed by more than one geofeed whose geofeeds disagree by the most severe difference.}
  \label{fig:multi_differences}
\end{figure}

To better understand duplicate geofeed entries, we now investigate the actual differences we see across these duplicates.
In total, we see 11.0k multi-entry prefixes with differences between multiple geofeeds out of the 280.7k (3.9\%) multi-entry prefixes.
Zooming in on the prefixes with differences as shown in \Cref{fig:multi_differences} (see \Cref{sec:appendix:sub:conflictrules} for details), we find that only 1.0k of these are format or naming variants (\eg GB vs. UK, JP-13 vs. TOKYO, Frankfurt vs. Frankfurt am Main).
With 6.2k, the largest share differ because one geofeed provides more detailed information compared to another (\eg city-level vs.\ country-level granularity, or GB-NET, Newcastle upon Tyne, vs.\ GB-ENG, England)---1.5k of these involve a geofeed from Larus~\cite{larusgeofeed}, an IP leasing company whose CEO also leads Cloud Innovation, the largest lessor of AFRINIC address space~\cite{du2024sublet} and infamous for its lawsuits against AFRINIC~\cite{beverly2026whereis,sharwood2026afrinic}.
The 6,747 prefixes in this geofeed are spread over 114 countries.
6,697 of these prefixes are from AFRINIC's address space, while only two AFRINIC \texttt{inetnum} objects reference this geofeed.
In contrast, Larus's other geofeed~\cite{larusgeofeedtrusted} (2.2k entries, 12 countries) is referenced by 372 WHOIS objects; both feeds share only 3 prefixes.

Continuing the analysis with substantive conflicts, we find 3.8k prefixes with such conflicts, \ie 1.4\% of all multi-entry prefixes:
1.4k prefixes with different city, 1.2k with different region, and another 1.2k with different country values.
Looking more closely at the substantive conflicts we find Larus emerging again as a large contributor with 429 country and 490 city conflicts.
For another 43 multi-entry prefixes, the locations differ only between entries within the same geofeed (\cf \S\ref{sec:inconsistencies:sub:within}).

\subsection{Authoritative Geofeeds for Conflicts}

As we find a non-negligible number of substantive conflicts in geofeeds, we now investigate which of these geofeeds is authoritative.
Following RFC~9632~\cite{rfc9632}, a geofeed is authoritative for a prefix if it is referenced by the most specific \texttt{inetnum} or \texttt{inet6num} object that covers the prefix and carries a geofeed reference.
If several such objects cover exactly the same address range, the most recently modified one is used.
For ARIN, we additionally count geofeed references on Org objects for all NetRanges of the Org.

We find that the vast majority of multi-entry prefixes without a substantive conflict (266.3k, 96.1\%) have exactly one authoritative geofeed.
For another 10.6k (3.8\%) of them, none of the listing geofeeds is authoritative: no covering \texttt{inetnum} references a geofeed, or the referenced geofeed could not be downloaded or does not list the prefix.

For substantive conflicts, we find a much lower share of 73.6\% (2793) of prefixes with exactly one authoritative geofeed, while the case without any authoritative geofeed is much more prevalent (990, 26.1\%).

Multiple geofeeds can be referenced at the most specific level for two reasons:
(1) an \texttt{inetnum} object carries multiple geofeed references (\eg two \texttt{Comment: Geofeed …} lines), or (2) several \texttt{inetnum} objects cover exactly the same address range, each with its own geofeed reference.
RFC~9632~\cite{rfc9632} resolves the second case by preferring the most recently modified object, which decides 28 multi-entry prefixes, 7 of them substantive conflicts.
For the first case, it only states that an object SHOULD have at most one geofeed reference, so all referenced geofeeds remain authoritative; this applies to 96 multi-entry prefixes, including 10 substantive conflicts.

Both cases stem mostly from a single operator, Bunny Communications, in the ARIN region.
One example prefix is 207.145.6.0/24, which has two NetRange objects covering exactly this /24: NET-207-145-6-0-1 (a reallocation to Bunny) and NET-207-145-6-0-2 (a reassignment to HGC Global Communications, with a parent link to NET-207-145-6-0-1).
The parent object references a single geofeed in a comment field~\cite{bunnycommunicationsgeofeed}.
The child object references two geofeeds, each in a comment field: the parent's geofeed and a different geofeed~\cite{bunnytechnologygeofeed}.
The child-only geofeed lists the location for this prefix as Sydney, Australia (\texttt{207.145.6.0/24,\allowbreak AU,\allowbreak AU-NSW,\allowbreak Sydney,}), whereas the shared geofeed lists it as Singapore (\texttt{207.145.6.0/24,\allowbreak SG,,\allowbreak Singapore,}).
As the parent was modified more recently (June 1, 2026) than the child (September 25, 2025), only the parent's geofeed is authoritative, even though the child is a delegation below the parent.
Even more concerningly, the child-only geofeed contradicts itself: it lists 13 multi-entry prefixes with more than one country, and for 207.145.6.0/24 it contains a malformed entry without prefix length (\texttt{207.145.6.0/,\allowbreak HK,\allowbreak HK-HCW,\allowbreak Hong Kong,}), most likely the Hong Kong entry for this /24, which RFC~8805 requires parsers to discard~\cite{rfc8805}.
The 10 substantive conflicts that keep multiple authoritative geofeeds are Bunny prefixes as well, whose NetRanges reference two of Bunny's geofeeds; these agree (Washington, DC), and only a non-authoritative geofeed disagrees (\texttt{US-WA} instead of \texttt{US-DC}).

Our findings paint a rather dim picture of geofeed data quality and highlight real-world issues that RFC~9632~\cite{rfc9632} does not resolve, such as an \texttt{inetnum} object referencing several geofeeds.

\subsection{Geographical Differences}

Next, we investigate whether there are substantial differences with regard to multi-entry prefixes and geofeed conflicts in the five RIR regions or specific countries.
We attribute each prefix to the RIR that manages its address space, based on the WHOIS dumps.
We see a wide range of multi-entry prefix shares, \ie the share of an RIR's geofeed prefixes that are listed by at least two geofeeds, ranging from low percentages of 0.8\% for LACNIC and 2.3\% for APNIC, to medium percentages for RIPE NCC with 8.8\%, and finally high percentages for AFRINIC and ARIN with 26.2\% and 28.5\%, respectively.
In terms of the share of substantive conflicts among multi-entry prefixes, we again see a large spread ranging from low shares for ARIN (0.6\%) and RIPE NCC (1.8\%) to high shares for APNIC (18.7\%) and AFRINIC (33.2\%).
However, few geofeeds drive these shares:
ARIN's low share is due to the large, agreeing geofeed pairs of Metronet and Cloudflare (see \Cref{sec:inconsistencies:sub:across}), 353 of APNIC's 473 conflicts stem from a single pair of geofeeds (as149841.net and virtutel.com.au), and 914 of AFRINIC's 929 conflicts involve Larus's geofeed with 6,747 prefixes, without which AFRINIC's share drops to 5.6\%.
We also see different shares of multi-entry prefixes without an authoritative geofeed per RIR: RIPE NCC (3.1\%) and ARIN (3.6\%) have low shares, APNIC a medium share (15.1\%), and AFRINIC a high share (56.9\%).
LACNIC has no substantive conflicts, but none of its 74 multi-entry prefixes has an authoritative geofeed: they are listed only by two copies of the same data (\eg Equinix's geofeed and a Google Sheets copy of it, or the opengeofeed.org aggregator~\cite{opengeofeed} and the geofeeds it hosts), and no WHOIS object covering them references a geofeed.
Of these prefixes, 48 are in Brazilian address space, where only 2 of the 223k \texttt{inetnum} and \texttt{inet6num} objects in the LACNIC WHOIS reference a geofeed.

\begin{figure}[t]
  \centering
  \includegraphics[width=\columnwidth]{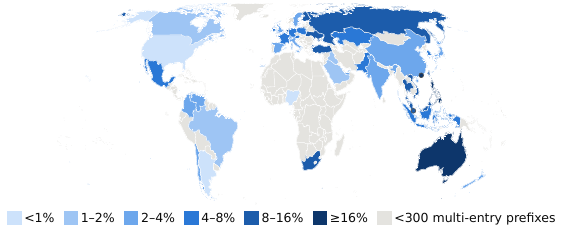}
  \caption{Share of multi-entry prefixes with a substantive conflict per country.}
  \label{fig:conflicts_by_country}
\end{figure}

Shifting our attention to differences per country, we show the share of multi-entry prefixes with a substantive conflict per country in \Cref{fig:conflicts_by_country}.
A prefix counts for every country that any of its geofeeds states, \eg a prefix that one geofeed locates in Germany and another in the Netherlands counts for both countries.
We consider the 45 countries with at least 300 multi-entry prefixes.
The highest share of conflicts can be found in Australia with 19.0\%: 629 of its 688 conflicts stem from disagreements between the geofeeds of as149841.net and virtutel.com.au; without this pair, Australia's share drops to 2.1\%.
Hong Kong (18.0\%), the Philippines (17.0\%), Malaysia (14.9\%), and Thailand (14.2\%) have high shares of conflicts largely due to Larus's geofeed with 6,747 prefixes (see \Cref{sec:inconsistencies:sub:conflicts}); without its prefixes, their shares drop to 7.7\%, 5.1\%, 2.2\%, and 1.4\%, respectively.
In contrast, the conflicts of Russia (14.2\%), the Netherlands (13.5\%), and Finland (10.4\%) are not caused by a few geofeeds: they are spread over 92, 140, and 39 geofeed pairs, and the shares remain high without each country's top geofeed pair (13.3\%, 12.5\%, and 8.8\%) and without Larus's geofeed (13.2\%, 12.1\%, and 10.5\%).
Most of these are country conflicts (112 of 198, 134 of 182, 38 of 47), mostly with neighboring countries or hosting hubs: RU with DE (13), NL and KZ (9 each); NL with DE (33), US (17) and CZ (10); FI with DE (7), NL (6), LV and US (5 each).
At the low end are countries dominated by large, agreeing geofeed pairs, such as the US (0.6\%), Canada (1.3\%), Brazil (1.4\%), and the UK (2.1\%).

 \section{Conclusion}

In this paper we analyzed geofeed errors, inconsistencies, and flaws, crawling the geofeeds referenced in the WHOIS of all five RIRs weekly over three months.
We find flaws at every level of the ecosystem:
First, WHOIS geofeed references often violate RFC~9632, \eg by using HTTP or listing multiple geofeeds per object; a single operator lowered compliance from 90\% to 61\%.
Second, 12--20\% of geofeed URLs cannot be downloaded each week.
Third, one in five geofeeds violates the RFC~8805 format, \eg with invalid region codes, and 17\% list IPv4 prefixes whose WHOIS object references another geofeed.
Fourth, 3.8k of the 281k prefixes listed in multiple geofeeds have conflicting locations, 26\% of them without an authoritative geofeed, and single geofeeds drive many of these conflicts.
Finally, RFC~9632~\cite{rfc9632} does not resolve real-world cases such as an inetnum object referencing multiple contradicting geofeeds.

These errors prevent a large fraction of geofeed entries from being
used as intended.  To improve the data quality of the geofeed
ecosystem, we developed a public web dashboard and shared it with the
network operator community. One large CDN used our dashboard to 
fix their geofeeds and shared that the root cause was due to 
multiple systems populating the WHOIS not being in sync.  Our hope is
that this work and the dashboard serve to aid network operators, 
geolocation providers, and
the geofeed ecosystem.
 
\clearpage
\bibliographystyle{ACM-Reference-Format}

\appendix
\section{Ethics}

All of the data we examine in this paper is publicly available and
does not include personally identifiable information.  In gathering
geofeeds, we minimize any potential impact on hosting infrastructure
by downloading any given geofeed at most once a week.
To validate our coverage methodology (\S\ref{sec:methodology}), we
publish a canary geofeed, referenced by our own inet6num objects, that
lists one IPv4 and one IPv6 prefix of a third-party network.  The
canary entries state exactly the location that the network's own,
authoritative geofeed gives for these prefixes, so consumers that
ingest them despite RFC~9632's coverage rule are not misled.

\section{Open Science}

\begin{figure*}
\centering
\includegraphics[width=0.9\linewidth]{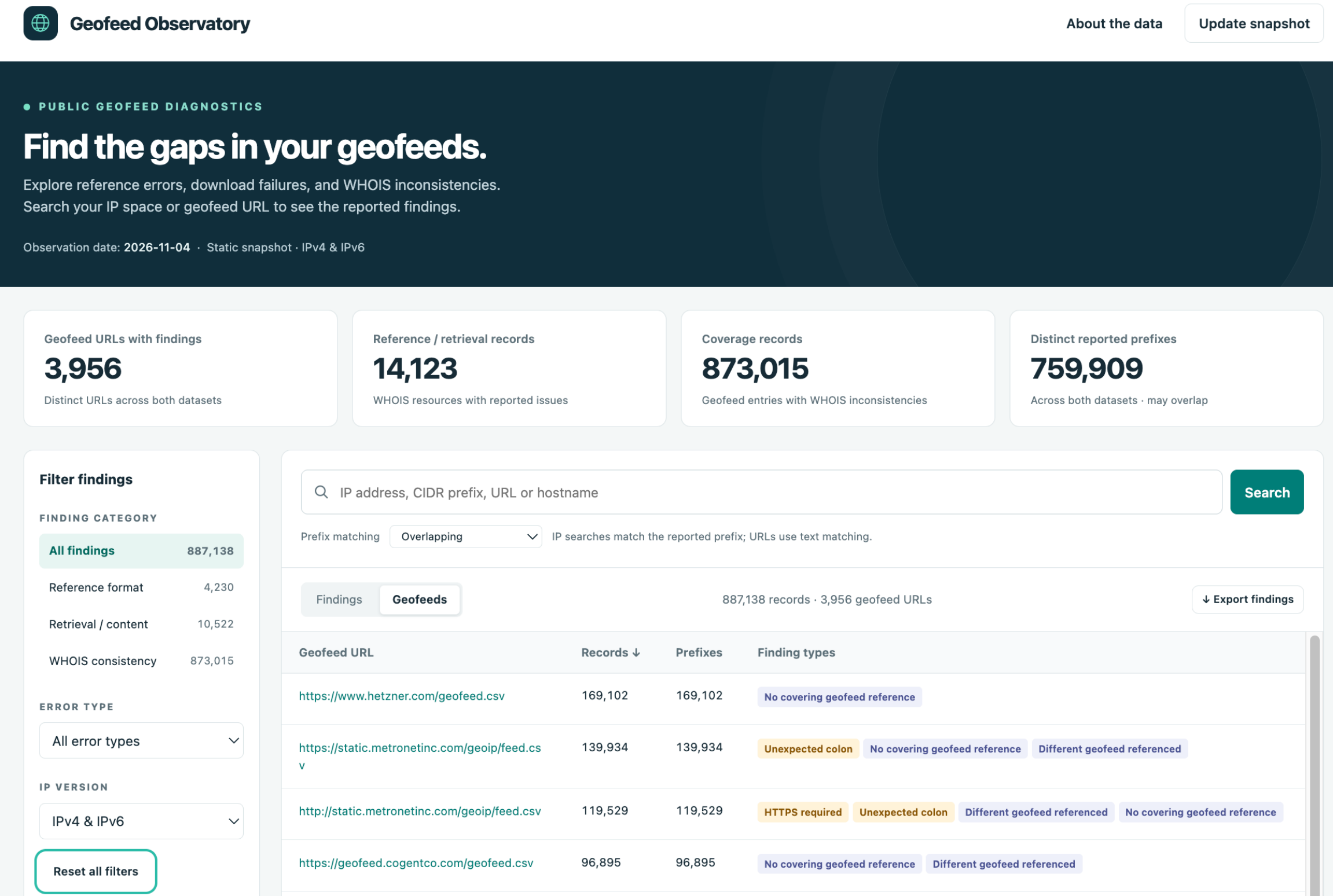}
\caption{Geoflaw Geofeed Observatory Dashboard provides an open
interface to query and download our results.}
\label{fig:dashboard}
\end{figure*}

We make our results publicly available via the Geoflaw
dashboard~\cite{geoflawweb}, a website that provides a queryable
interface to the errors and inconsistencies we discover, as well as a
means to download the raw data in CSV form.  A screenshot of the
dashboard is included here as Figure~\ref{fig:dashboard} for reference.

\section{Generative AI}

We used generative AI to find typos and to correct grammatical errors
and awkward phrasing throughout the paper.  We use
OpenAI GPT-6 Astra and local Gemma-4.5 models to create the web
dashboard~\cite{geoflawweb} and to produce Python code that robustly
downloads geofeed files.  The remainder of the analysis scripts and
code are created by humans.

\section{Conflict Rules}
\label{sec:appendix:sub:conflictrules}

Different conflict rules are shown in \Cref{tab:classification}.

\begin{table*}[t]
  \centering
  \footnotesize
  \caption{Classification of the multi-entry prefixes, \ie prefixes listed by at least two geofeeds, by how the geofeeds' locations for them differ. Each geofeed contributes the entries of its representative URL; we compare country, region, and city, and ignore postal codes. Regions are resolved with ISO~3166-2 as distributed in Debian's \texttt{iso-codes} 4.20.1. Examples show normalized values as (country, region, city); -- is an empty field.}
  \label{tab:classification}
  \begin{tabularx}{\textwidth}{@{}>{\RaggedRight\arraybackslash}p{3.2cm}>{\RaggedRight\arraybackslash}X>{\RaggedRight\arraybackslash}p{5.3cm}r@{}}
    \toprule
    & Rule & Example & Prefixes \\
    \midrule
    \multicolumn{4}{@{}>{\RaggedRight\arraybackslash}p{\textwidth}@{}}{\textbf{Step 1: Normalization} of every entry. Prefixes whose raw values differ only in these respects count as identical (277 prefixes).} \\
    Country & Trim whitespace, upper-case & \texttt{us} $\rightarrow$ \texttt{US} & \\
    Region & Trim whitespace, upper-case; empty if equal to the country code & \texttt{Tokyo} $\rightarrow$ \texttt{TOKYO}; \texttt{HK} in \texttt{HK} $\rightarrow$ -- & \\
    City & Collapse and trim whitespace, strip accents, lower-case & \texttt{MONTR\'EAL} $\rightarrow$ \texttt{montreal} & \\
    \midrule
    \multicolumn{4}{@{}>{\RaggedRight\arraybackslash}p{\textwidth}@{}}{\textbf{Step 2: Region resolution.} Each region value is mapped to ISO~3166-2 subdivisions of its entry's country by the first matching rule. Names are compared to the ISO names and their English translations, ignoring case, accents, and punctuation.} \\
    ISO code & The value is an ISO~3166-2 code & \texttt{JP-13} $\rightarrow$ \texttt{JP-13} & \\
    Whole country & The value is \texttt{CC-CC} or the country's name & \texttt{FRANCE} $\rightarrow$ \texttt{FR} & \\
    Bare code & With the country code prefixed, it is an ISO~3166-2 code & \texttt{CA} in \texttt{US} $\rightarrow$ \texttt{US-CA} & \\
    Name & Equal to the name of a subdivision & \texttt{TOKYO} $\rightarrow$ \texttt{JP-13} & \\
    Name prefix & One name starts with the other (at least 4 characters) & \texttt{STOCKHOLM} $\rightarrow$ \texttt{SE-AB} (Stockholms län) & \\
    Similar name & Most similar name, with \mbox{Jaro-Winkler} similarity $\geq 0.85$ & \texttt{SAO PAOLO} $\rightarrow$ \texttt{BR-SP} (São Paulo) & \\
    Unresolved & No rule matches, \eg obsolete or unofficial codes & \texttt{FR-75} (Paris before 2019), \texttt{RU-SPB} & \\
    \midrule
    \multicolumn{4}{@{}>{\RaggedRight\arraybackslash}p{\textwidth}@{}}{\textbf{Step 3: Value pairs.} For each prefix and field, every pair of distinct non-empty values $a$, $b$ where one geofeed states $a$ but not $b$ and another $b$ but not $a$ gets the first matching class. Per field, a prefix counts once, under the most severe class of its pairs (different $>$ unresolved $>$ nested $>$ same subdivision, naming variant $>$ format, alias); it can count in several fields.} \\
    \emph{Country} & & & \\
    \quad Alias & \texttt{UK} and \texttt{GB} & \texttt{UK} / \texttt{GB} & 13 \\
    \quad \textbf{Different} & Any other pair & \texttt{TH} / \texttt{US} & 1249 \\
    \emph{Region} & & & \\
    \quad Format & Equal without the \texttt{CC-} prefix and non-alphanumeric characters & \texttt{US-CA} / \texttt{CA} & 46 \\
    \quad Same subdivision & Both resolve to the same subdivision & \texttt{JP-13} / \texttt{TOKYO} & 343 \\
    \quad Nested & One lies within the other (ISO~3166-2 parents), or is the whole country & \texttt{GB-ENG} / \texttt{GB-NET} & 3585 \\
    \quad \textbf{Unresolved} & At least one value is unresolved; counted as different & \texttt{FR-75} / \texttt{FR-IDF} & 180 \\
    \quad \textbf{Different} & Any other pair & \texttt{DE-HE} / \texttt{DE-HH} & 1557 \\
    \emph{City} & & & \\
    \quad Format & Equal after replacing punctuation by spaces & \texttt{hong kong} / \texttt{hong-kong} & 5 \\
    \quad Naming variant & \mbox{Jaro-Winkler} similarity $\geq 0.85$, or one name contains the other & \texttt{warsaw} / \texttt{warszawa} & 949 \\
    \quad \textbf{Different} & Any other pair & \texttt{santa clara} / \texttt{abbeville} & 3494 \\
    \midrule
    \multicolumn{4}{@{}>{\RaggedRight\arraybackslash}p{\textwidth}@{}}{\textbf{Step 4: Prefix outcome.} Each prefix counts once, under the first matching outcome. The first three are substantive conflicts.} \\
    \textbf{Different country} & A country pair is \emph{different} & (TH, TH-10, bangkok) vs.\newline (US, --, abbeville) & 1249 \\
    \textbf{Different region} & A region pair is \emph{different} or \emph{unresolved} & (DE, DE-HE, frankfurt/main) vs.\newline (DE, DE-HH, hamburg) & 1176 \\
    \textbf{Different city} & A city pair is \emph{different} & (US, --, abbeville) vs.\newline (US, US-CA, santa clara) & 1368 \\
    Format / naming variant & At least one value pair other than \emph{nested}, none of them \emph{different} or \emph{unresolved}; may additionally differ in detail & (DE, DE-HE, frankfurt/main) vs.\newline (DE, HESSE, frankfurt) & 1020 \\
    More vs.\ less detail & No value pair other than \emph{nested}; the locations differ only because one geofeed leaves a field empty that another fills, or states a region that contains the other's & (DE, --, berlin) vs.\newline (DE, DE-BE, berlin); (GB, GB-ENG, newcastle upon tyne) vs.\newline (GB, GB-NET, newcastle upon tyne) & 6214 \\
    \cmidrule(l){2-4}
    & \multicolumn{2}{l}{Prefixes whose locations differ between geofeeds} & 11027 \\
    \addlinespace
    Inconsistent within a \mbox{geofeed} & Values differ only between entries of the same geofeed, which lists the prefix several times & (HK, HK-HCW, hong kong) and\newline (SG, --, singapore) in one geofeed & 43 \\
    Identical & All geofeeds give the same location & & 269657 \\
    \cmidrule(l){2-4}
    & \multicolumn{2}{l}{Multi-entry prefixes} & 280727 \\
    \bottomrule
  \end{tabularx}
\end{table*}
 
\label{page:last}

\end{document}